\documentclass[%
 reprint,
superscriptaddress,
amsmath,amssymb,
aps,
prb,
longbibliography]{revtex4-2}

\usepackage{graphicx}
\usepackage{dcolumn}
\usepackage{bm}%
\usepackage{graphicx,graphics,color,epsfig}
\usepackage{amsmath}
\usepackage{mathrsfs}
\usepackage{amssymb}
\usepackage{subfigure}
\usepackage{amsfonts}
\usepackage{graphicx}
\usepackage{multirow,booktabs}
\usepackage{float}
\usepackage{natbib}
\usepackage[normalem]{ulem}  
\usepackage{xcolor}
\usepackage{booktabs}    
\usepackage{array}

\begin{document}

\title{The emergence of long-range entanglement and odd-even effect in periodic generalized quantum cluster models}

\author{Zhen-Yu Zheng}
    \affiliation{Beijing National Laboratory for Condensed Matter Physics, Institute
of Physics, Chinese Academy of Sciences, Beijing 100190, China}

\author{Shu Chen}
\thanks{Corresponding author: schen@iphy.ac.cn}
\affiliation{Beijing National Laboratory for Condensed Matter Physics, Institute
of Physics, Chinese Academy of Sciences, Beijing 100190, China}
\affiliation{School of Physical Sciences, University of Chinese Academy of Sciences,
Beijing 100049, China}
\date{\today}

\begin{abstract}
We investigate the entanglement properties in a generalized quantum cluster model under periodic boundary condition. By evaluating the quantum conditional mutual information entropy under four subsystem partitions, we identify clear signatures of  long-range entanglement. Specifically, when both  the system size $N$ and the interaction range $m$ are odd, the system exhibits nonzero four-part quantum conditional mutual information entropies in infinitesimal but finite field. This non-vanishing four-part quantum conditional mutual information entropy directly signals the presence of long-range entanglement. In contrast, all other
combination of $N$ and $m$ yield vanishing four-part quantum conditional mutual information entropy. Remarkably, in the case of $N, m \in \text{odd}$, these long-range entangled features persist even in the presence of a large transverse field, demonstrating their robustness against quantum fluctuations. These results demonstrate how the interplay between system size  and interaction range governs the emergence of long-range entanglement in one-dimensional generalized quantum cluster model.
\end{abstract}

\maketitle

\section{Introduction}
Quantum entanglement has emerged as an indispensable diagnostic for characterizing and classifying the rich phases of many-body quantum systems\cite{Amico_2008_RMP,Calabrese_2009_JPA,Eisert_2010_RMP,Fradkin_2013_field}. A typical example is the topological entanglement entropy (TEE) in two-dimensional systems, which unambiguously detects intrinsic topological order\cite{HAMMA_2005_PLA,Levin_2006_PRL,Kitaev_2006_PRL}. By isolating a universal sub-leading constant in the entanglement scaling, the TEE quantifies the long-range entanglement pattern that is not dis-tillable by local operations, thereby serving as a genuine order parameter that distinguishes phases based on their quasi-particle content\cite{Wen_2017_RMP,Wen_2019_Sci}.

However, applying the TEE to one-dimensional systems presents a significant challenge. In one-dimensional, standard bipartite entanglement measures of connected partitions fail to act as universal order parameters\cite{Calabrese_2009_JPA,Eisert_2010_RMP,Alioscia_2005_PRA}. Their scaling lacks informative corrections, and their structure is strongly influenced by non-universal, ultraviolet contributions from boundaries. Even finer metrics like the entanglement spectrum cannot reliably distinguish the topological character of different wave functions\cite{Li_2008_PRL,Pollmann_2010_PRB,Fidkowski_2010_PRL,Turner_2011_PRB}.
To overcome these limitations and isolate the universal information locked in non-local correlations, multi-partite measures are required. A powerful such tool is the four-part conditional mutual information\cite{Zeng_2019_quantum}, also known as the disconnected entanglement entropy (DEE)\cite{Fromholz_2020_PRB,Micallo_2020_scicore,Torre_2024_scicore}. By its very construction, the DEE is designed to systematically cancel the problematic local and boundary-driven entanglement contributions. Consequently, a non-zero DEE value serves as a robust indicator of the genuine long-range correlations and long-range entanglement that define topological phases in 1D\cite{Torre_2024_scicore}, a theoretical utility that is complemented by its prospective relevance for experiments\cite{Hannes_2016_PRX,Kenny_2018_PRL,Vermersch_2018_PRA,Lavasani_2021_NP}.

Cluster models, also known as the $\alpha$-chain, and their generalizations possess rich topological structures and well-defined order parameters, making them ideal platforms for exploring the interplay between topology and quantum correlations\cite{Pachos_2004_PRL,Doherty_2009_PRL,Smacchia_2011_PRA,Niu_2012_PRB,Montes_2012_PRE,Son_2012_QIP,
DeGottardi_2013_PRB,Lahtinen_2015_PRL,Zhang_2015_PRL,Giampaolo_2015_PRA,Zeng_2016_EL,Hannes_2016_PRX,Potirniche_2017_PRL,
Verresen_2017_PRB,Nie_2017_PRE,Scaffidi_2017_PRX,Zhang_2018_PRL,Bhattacharjee_2018_PRB,Zonzo_2018_JSM,Ding_2019_PRE,Guo_2022_PRA,Fabio_2022_PRB,Li_2025_PRA,
Yi_2025_PRA}.
They provide a paradigmatic framework for investigating one-dimensional symmetry-protected topological (SPT) phases.
Owing to their stabilizer-type construction, these models not only capture the essential features of SPT order in spin chains but also find extensive applications in quantum information processing, where cluster states serve as fundamental resources for measurement-based quantum computation\cite{Kenny_2018_PRL,Lavasani_2021_NP,Ippoliti_2021_PRX,Morral_2023_PRB,Qian_2024_PRB,Yu_2025_PS,Bera_2025_arxiv}.
The ground states of such models exhibit nonlocal stabilizer correlations and characteristic topological degeneracy, while the inclusion of a transverse field drives a quantum phase transition between the topological cluster phase and a trivial paramagnetic phase.
However, existing studies of their entanglement properties have been predominantly restricted to systems with open boundary condition (OBC). Although the ground states of generalized cluster models under OBC exhibit highly non-trivial properties which present as SPT order or mixed order (having SPT order and symmetry-breaking order)\cite{Zeng_2019_quantum},  the entanglement in all such instances remains confined to the short-range entanglement, with no long-range entanglement observed. Meanwhile, the PBC counterpart is also commonly regarded as incapable of supporting long-range entanglement and the entanglement structure of generalized cluster models under PBC has received far less attention.

In this work, we investigate the entanglement properties of generalized quantum cluster models under PBC and demonstrate that long-range entanglement can indeed persist under PBC when both the system size $N$ and interaction range $m$ are odd. By analyzing the entanglement entropy and the multi-part conditional mutual information entropy, we identify clear signatures of long-range entanglement in this regime, even in the presence of a finite transverse field. Our findings reveal that the size of the system and the interaction range play a crucial role in protecting non-local correlations under PBC. Furthermore, by combining spectral and entanglement analyses, we show that this mechanism leads to gapless low-energy modes and robust long-range entanglement characteristics.

The remainder of this paper is organized as follows: Section~\ref{sec:model} introduces the generalized quantum cluster model and the ground states are exactly solved in different system size and interaction range conditions with zero-th order approximation. Section~\ref{sec:entanglement} presents numerical results for  quantum condition mutual information across different conditions in infinitesimal but finite field. Section~\ref{sec:robustness} we investigate the entanglement properties under a large quantum fluctuation to prove the robust of long-range entanglement. Finally, Section~\ref{sec:conclusion} summarizes our findings and outlines future directions.

\section{Model and exact solution of ground-state}\label{sec:model}
We investigate a family of one-dimensional generalized quantum cluster models characterized by multi-spin interaction Hamiltonian of the form:
\begin{equation}
  H = J \sum_{j = 1}^{N} \sigma_{j}^{x} \tau_{j, m}^{z} \sigma_{j + m}^{x} + H_{z},
  \label{Hm}
\end{equation}
where the cluster operator $\tau_{j, m}^{z} = \sigma_{j + 1}^{z} \cdots \sigma_{j + m - 1}^{z}$
spans $m - 1$ intermediate sites, with $m$ determining the interaction range of each multi-spin term. The Pauli matrices $\sigma_i^\alpha$ ($\alpha = x, z$) act on site $i$ and $H_{z}=-h \sum_{j=1}^{N}\sigma_{j}^{z}$ is a term due to transverse field which can be referred to the quantum fluctuation. For convenience, assume the interaction strength $J = 1$ and periodic boundary conditions ($\sigma_{N+1} = \sigma_{1}$) throughout our analysis. We first consider the the zero-field case as a zeroth-order approximation ($h=0$), which allows for an analytical treatment. Using the Jordan–Wigner transformation,
\begin{equation}\label{JWT}
\begin{aligned}
\sigma_{j}^{x} &= (c_{j}^{\dagger} + c_{j}) \mathrm{e}^{\mathrm{i}\pi \sum_{l<j} c_{l}^{\dagger} c_{l}}, \quad
\sigma_{j}^{z} = 2 c_{j}^{\dagger} c_{j} -1 ,
\end{aligned}
\end{equation}
the Hamiltonian can be rewritten as
\begin{equation}\label{eq:pz}
H = P_{z}^{-} H^{+} P_{z}^{-} + P_{z}^{+} H^{-} P_{z}^{+},
\end{equation}
where the projectors are given by
\begin{equation}
P_{z}^{\pm}=\tfrac{1}{2}(1\pm\mathcal{P}_{z}), \quad
\mathcal{P}_{z}=\mathrm{e}^{\mathrm{i} \pi \sum_{j} c_{j}^{\dagger} c_{j}} .
\end{equation}
Here $\mathcal{P}_{z}$ denotes the fermion-parity operator. The two fermion Hamiltonians take the form,
\begin{equation}
\begin{aligned}
H^{\pm} &= (-1)^{m-1} \Biggl[
\sum_{j=1}^{N-m} \bigl(c_{j}^{\dagger} - c_{j}\bigr)\bigl(c_{j+m}^{\dagger} + c_{j+m}\bigr) \\
& - \mathcal{P}_{z} \sum_{i=1}^{m} \bigl(c_{N-m+i}^{\dagger} - c_{N-m+i}\bigr)\bigl(c_{i}^{\dagger} + c_{i}\bigr)
\Biggr].
\end{aligned}
\end{equation}
It follows that $H^{+}$ corresponds to periodic boundary conditions (PBC) with $c_{N+1}=c_{1}$, while $H^{-}$ corresponds to anti-periodic boundary conditions (APBC) with $c_{N+1}=-c_{1}$.

We first consider system with odd size ($N \in$ odd). Under PBC and APBC, the quantized momentum values are respectively given by:
\begin{equation}
\begin{gathered}
Q^{+} \in \left\{ -\tfrac{N-1}{N}\pi,\, \ldots,\, -\tfrac{2}{N}\pi,\, 0,\, \tfrac{2}{N}\pi,\, \ldots,\, \tfrac{N-1}{N}\pi \right\}, \\
Q^{-} \in \left\{ -\tfrac{N-2}{N}\pi,\, \ldots,\, -\tfrac{1}{N}\pi,\, \tfrac{1}{N}\pi,\, \ldots,\, \tfrac{N-2}{N}\pi,\, \pi \right\},
\end{gathered}
\end{equation}
where $Q^{+}$ and $Q^{-}$ correspond to PBC and APBC cases.

By applying the Fourier transform
$c_{j}^{\dagger} = \tfrac{1}{\sqrt{N}} \sum_{q} c_{q}^{\dagger}\, \mathrm{e}^{\mathrm{i} q j}$
together with the Bogoliubov transformation
$\eta_{q} = u_{q} c_{q} - i v_{q} c_{-q}^{\dagger}$,
the two fermionic Hamiltonians in momentum space can be diagonalized as
\begin{equation}
\begin{gathered}
H^{+} = \sum_{q \in q^{+}} \left(2\eta_{q}^{\dagger}\eta_{q} - 1\right)
- (-1)^{m} \left(2c_{0}^{\dagger}c_{0} - 1\right), \\
H^{-} = \sum_{q \in q^{-}}  \left(2\eta_{q}^{\dagger}\eta_{q} - 1\right)
- \left(2c_{\pi}^{\dagger}c_{\pi} - 1\right),
\end{gathered}
\end{equation}
where the allowed momenta are restricted to
$q^{+} = \left\{\tfrac{2\pi m}{N}\,\big|\, m=1,2,\dots,\tfrac{N-1}{2}\right\}$
and
$q^{-} = \left\{\tfrac{2\pi m-1}{N}\,\big|\, m=1,2,\dots,\tfrac{N-1}{2}\right\}$, respectively.
The Bogoliubov coefficients are given by
\begin{equation}
\begin{gathered}
u_{q}^{2} = \tfrac{1}{2}\bigr(1 - (-1)^{m} \cos(mq)\bigr);
\\v_{q}^{2} = \tfrac{1}{2}\bigr(1 + (-1)^{m} \cos(mq)\bigr);
\\ u_{q} v_{q} = -(-1)^{m}  \sin(mq), \\
\end{gathered}
\end{equation}

For clarity in subsequent discussions, we denote the vacuum states of $H_{m}^{+}$ and $H_{m}^{-}$ in $\eta$ representation as $|\varphi^{+}\rangle$ and $|\varphi^{-}\rangle$, respectively. Their explicit expressions read:
\begin{equation}
\begin{gathered}
|\varphi^{+}\rangle = \prod_{q \in Q^{+}}\left(u_{q}+ i v_{q} c_{q}^{\dagger}c_{-q}^{\dagger}\right)|0\rangle,\\
|\varphi^{-}\rangle = \prod_{q \in Q^{-}}\left(u_{q}+ i v_{q} c_{q}^{\dagger}c_{-q}^{\dagger}\right)|0\rangle,
\end{gathered}
\end{equation}
where $|0\rangle$ denotes the fermionic vacuum.

Based on Eq.(\ref{eq:pz}), we find that only states with an odd number of occupied fermions are allowed for $H^{+}$, while only those with an even number of occupied fermions are valid for $H^{-}$
. All other states are redundant for the full Hamiltonian $H$. Consequently, the ground-state degeneracy structure is fundamentally determined by the range $m$ of the multi-spin interaction, revealing two distinct scenarios:
(i) For even $m$: The system presents unique ground states: $c_0^\dag|\varphi^+\rangle$; (ii) For odd $m$: The degeneracy of ground state expands to $2N$ and ground states include:
$c_0^\dag|\varphi^+\rangle$, $\eta_{q\in Q^{+}, q\neq 0}^{\dag}|\varphi^+\rangle$,
$\eta_{q\in Q^{-}, q\neq \pi}^{\dag}\eta_{\pi}^{\dag}|\varphi^-\rangle$, and $|\varphi^-\rangle$. This degeneracy exhibits system size dependent scaling.

In contrast, for even system size ($N \in$ even), the quantized momentum values for PBC and APBC are given by:
\begin{equation}
\begin{gathered}
Q^{+} \in \left\{ -\tfrac{N-2}{N}\pi,\, \ldots,\, -\tfrac{2}{N}\pi,\, 0,\, \tfrac{2}{N}\pi,\, \ldots,\, \tfrac{N-2}{N}\pi, \pi \right\}, \\
Q^{-} \in \left\{ -\tfrac{N-1}{N}\pi,\, \ldots,\, -\tfrac{1}{N}\pi,\, \tfrac{1}{N}\pi,\, \ldots,\, \tfrac{N-1}{N}\pi \right\},
\end{gathered}
\end{equation}
and the following fermion Hamiltonians can be diagonalized as:
\begin{equation}
\begin{gathered}
H_m^{+} = \sum_{q\in q^+} \bigl(2\eta_{q}^{\dagger}\eta_{q}-1\bigr)-(-1)^{m} \bigl(2c_{0}^{\dagger}c_{0}-1\bigr)- \bigl(2c_{\pi}^{\dagger}c_{\pi}-1\bigr);
\\H_m^{-} = \sum_{q\in q^-} \bigl(2\eta_{q}^{\dagger}\eta_{q}-1\bigr).
\end{gathered}
\end{equation}
where the allowed momenta are restricted to
$q^{+} = \left\{\tfrac{2\pi m}{N}\,\big|\, m=1,2,\dots,\tfrac{N-2}{2}\right\}$
and
$q^{-} = \left\{\tfrac{2\pi m-1}{N}\,\big|\, m=1,2,\dots,\tfrac{N}{2}\right\}$, respectively. And the vacuum states of $H_{m}^{+}$ and $H_{m}^{-}$ in $\eta$ representation are:
\begin{equation}
\begin{gathered}
|\varphi^{+}\rangle = \prod_{q\in q^+}(u_{q}+\mathrm{i}v_{q}c_{q}^{\dagger}c_{-q}^{\dagger})|0\rangle,\\
|\varphi^{-}\rangle = \prod_{q\in q^-}(u_{q}+\mathrm{i}v_{q}c_{q}^{\dagger}c_{-q}^{\dagger})|0\rangle.
\end{gathered}
\end{equation}
We also find that the ground-state degeneracy depends on the range $m$ of the multi-spin interaction. For even $m$, the system hosts a unique ground state $|\varphi^{-}\rangle$. Conversely, for odd $m$, the ground state is doubly degenerate, consisting of $c_{\pi}^{\dagger}|\varphi^{+}\rangle$ and $|\varphi^{-}\rangle$.

The analysis of the ground-state degeneracy can also be carried out by mapping the model to another spin representation via the Kramers--Wannier transformation, as shown in the Appendix~\ref{appendixKWT}.

\section{Nonzero four-part quantum conditional mutual information }\label{sec:entanglement}
\begin{figure}[t]
  \centering
  \includegraphics[width=\columnwidth]{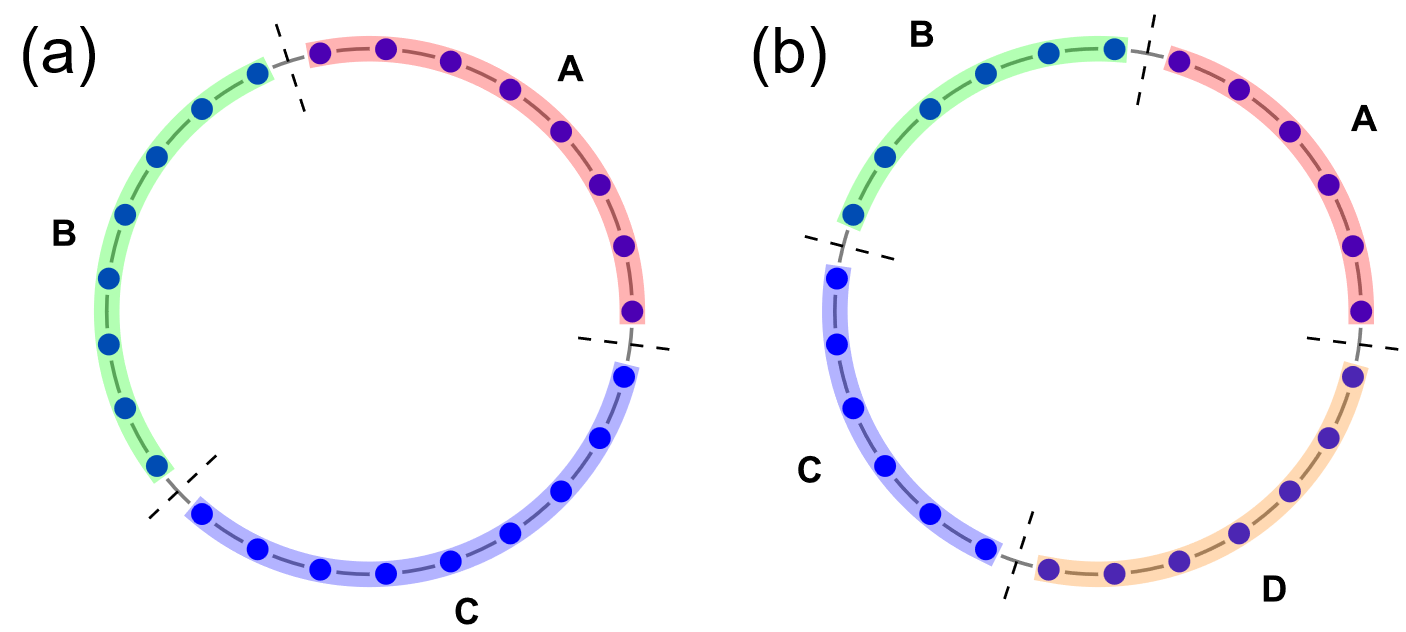}
  \caption{Schematic representation of the two types of partitions used in the calculation of the quantum mutual information entropy. (a) The system is divided into three subsystems, with the resulting entropy denoted by $S_{cmi}^{t}$. (b) The system is divided into four parts, with the corresponding entropy denoted by $S_{cmi}^{q}$.}
  \label{figsyt}
\end{figure}

To gain deeper insight into the entanglement properties of  generalized quantum cluster models, we main analyze the quantum conditional mutual information\cite{Zeng_2019_quantum,Torre_2024_scicore} to characterize their the nature of their long-range entanglement.

Notably, throughout this work we evaluate entanglement measures on pure ground states with well-defined parity and momentum. We first analyze the zero-field limit as a zeroth-order approximation, where analytical results can be obtained. In this limit, special equal-weight superpositions or statistical mixtures of all degenerate ground states may indeed reduce or even eliminate the entanglement.

To avoid ambiguities associated with ground-state degeneracy, we emphasize that once an infinitesimal but finite symmetry-preserving perturbation (such as a transverse field) is introduced, the lifting of the degeneracy is expected to be the leading qualitative effect. The system then selects a symmetry-resolved ground state with well-defined parity and momentum, rather than a special superposition or an incoherent mixture of degenerate ground states. Therefore, the symmetry-resolved pure ground states are the physically relevant states under such perturbations, and all entanglement results in this work should be understood in this sense. In particular, in the presence of an infinitesimal symmetry-preserving field, we focus on the symmetry-resolved ground states selected by this leading effect, such as $c_0^\dagger|\varphi^+\rangle$ and $|\varphi^-\rangle$. Therefore, all subsequent discussions are based on an infinitesimal but finite transverse field.

To investigate the nontrivial entanglement properties of the generalized quantum cluster model, we compute the quantum conditional mutual information $ S_{cmi} $ \cite{Zeng_2019_quantum,Wyner_1978_IC}, defined as
\begin{equation}\label{eq:Scmi}
    S_{cmi} = S_{AB} + S_{BC} - S_{B} - S_{ABC},
\end{equation}
where $S_{AB}$, $S_{BC}$, $S_{B}$, and $S_{ABC}$ denote the entanglement entropies $S_l$ of the corresponding subsystems, evaluated for different block sizes. For brevity, the definitions, analytical derivations, final expressions, and a more detailed discussion are shown in  Appendix~\ref{appendixEE}.

As illustrated in Fig.~\ref{figsyt}, we consider two types of partitions.
In panel (a), the system is divided into three subsystems, and the conditional quantum mutual information entropy is denoted as $ S_{\mathrm{cmi}}^{t} $.
In panel (b), the system is divided into four parts, and the corresponding entropy is denoted as $ S_{\mathrm{cmi}}^{q} $.
Although the partition is not unique in a general sense, we choose to divide the system as evenly as possible to simplify the calculation while ensuring this choice does not affect the physical conclusions.
For instance, when the total number of lattice sites is $N = 25$, we partition the system into three parts with subsystem sizes $A$, $B$, and $C$ containing 8, 8, and 9 sites, respectively.
Similarly, for the four-part partition, the subsystems $A$, $B$, $C$, and $D$ contain 6, 6, 6, and 7 sites, respectively.

When the system is partitioned into three subsystems, the conditional quantum mutual information captures only partial information. Therefore, by itself it cannot serve as an unambiguous diagnostic of long-range entanglement, although it can still reveal the presence of long-range correlations. We do not discuss this point in detail in the main text, and show a more detailed analysis in Appendix \ref{appendixTPMIC}.

\begin{figure}[t]
  \centering
  \includegraphics[width=0.85\columnwidth]{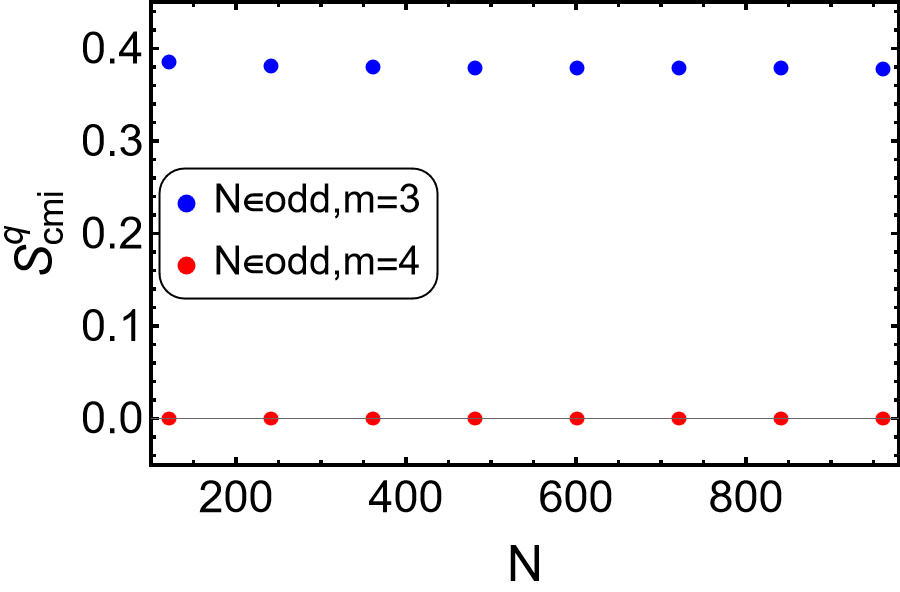}
  \caption{Four-part quantum mutual information entropy as a function of system size. Results are shown for two representative cases: $m=3$ with odd $N$ (blue points), where $S_{cmi}^{q}$ is nonzero, and $m=4$ with odd $N$ (red points), where $S_{cmi}^{q}$ vanishes.}
  \label{figCMIq}
\end{figure}

This limitation motivates us to introduce a more stringent diagnostic that can distinguish genuine long-range entanglement from merely long-range correlations. To isolate a purely topological invariant, we now turn to the disconnected entanglement entropy $S^D$, introduced in Ref.~\cite{Zeng_2019_quantum}.
Analogous to the topological entanglement entropy\cite{Kitaev_2006_PRL,Fendley_2007_JSP}, $S^D$ is constructed from a specific linear combination of entanglement entropies designed to cancel all non-universal contributions, thereby isolating a constant topological value.
The key distinction lies in the partitioning geometry, which allows $S^D$ to serve as a diagnostic for a broader class of topological phases, including both intrinsic topological orders and symmetry-protected topological phases, as validated previous works~\cite{Zeng_2019_quantum,Fromholz_2020_PRB,Micallo_2020_scicore,Wang_2018}.

The general definition of the disconnected entropy for a chain involving two disjoint intervals is given by Refs.~\cite{Casini_2004_PLB,Fromholz_2020_PRB,Micallo_2020_scicore,Torre_2024_scicore} as
\begin{equation}
    S^D = S_X + S_Y - S_{X \cup Y} - S_{X\cap Y},
    \label{eq:SD_general}
\end{equation}
where the complement $\overline{X \cup Y}$ and the intersection $X \cap Y$ are non-empty. Here, $S_l$ denotes the entanglement entropy.

For our analysis, we adopt the specific four-region partitioning scheme for the chain as illustrated in Fig.~1(b). In this scheme, we define the regions as $X = A \cup B$ and $Y = B \cup C$. Consequently, the intersection is $X \cap Y = B$, and the disconnected subset corresponds to $D = \overline{X \cup Y} = \overline{A \cup B \cup C}$. For such a configuration, the general expression in Eq.~(\ref{eq:SD_general}) simplifies significantly. Specifically, the combination of terms rearranges to become
\begin{equation}
    S^D = S_{AB} + S_{BC} - S_{B} - S_{ABC},
    \label{eq:SD_simplified}
\end{equation}
which is precisely the mutual information, $S_{cmi}^{q}$, defined in Eq.~(\ref{eq:Scmi}). Therefore, for our chosen configuration, the two quantities are equivalent, and we can use $S_{cmi}^{q}$ as our diagnostic. This measure effectively cancels local contributions to highlight the long-range entanglement characteristic of topological phases, thereby serving as a sensitive probe for their emergence.

Here we consider the system-size dependence for two representative classes of cases, namely, $m \in \text{odd},\, N \in \text{odd}$ and all the other cases. Since the entanglement entropy analytical analysis which presented in Appendix~\ref{appendixEE} has already shown that all cases exhibit similar behavior except when both $m$ and $N$ are odd, we focus on the case $m \in \text{even},\, N \in \text{odd}$ as a typical example.

As shown in Fig.\ref{figCMIq}, when $ m \in \text{odd} $ and $ N \in \text{odd} $, we observe a nonzero four-part quantum conditional mutual information entropy. This nonzero value maybe originates from the non-local nature of entanglement and signals long-range entanglement. The essential feature persists non-local entanglement leads to long-range correlation and entanglement. In contrast, for all other cases such as $ m \in \text{even}, N \in \text{odd} $, the four-part quantum mutual information entropy vanishes as shown in Fig.\ref{figCMIq}, indicating the absence of long-range entanglement.

\section{Robustness of the findings against a large field}\label{sec:robustness}
\begin{figure}[t]
  \centering
  \includegraphics[width=8.5cm]{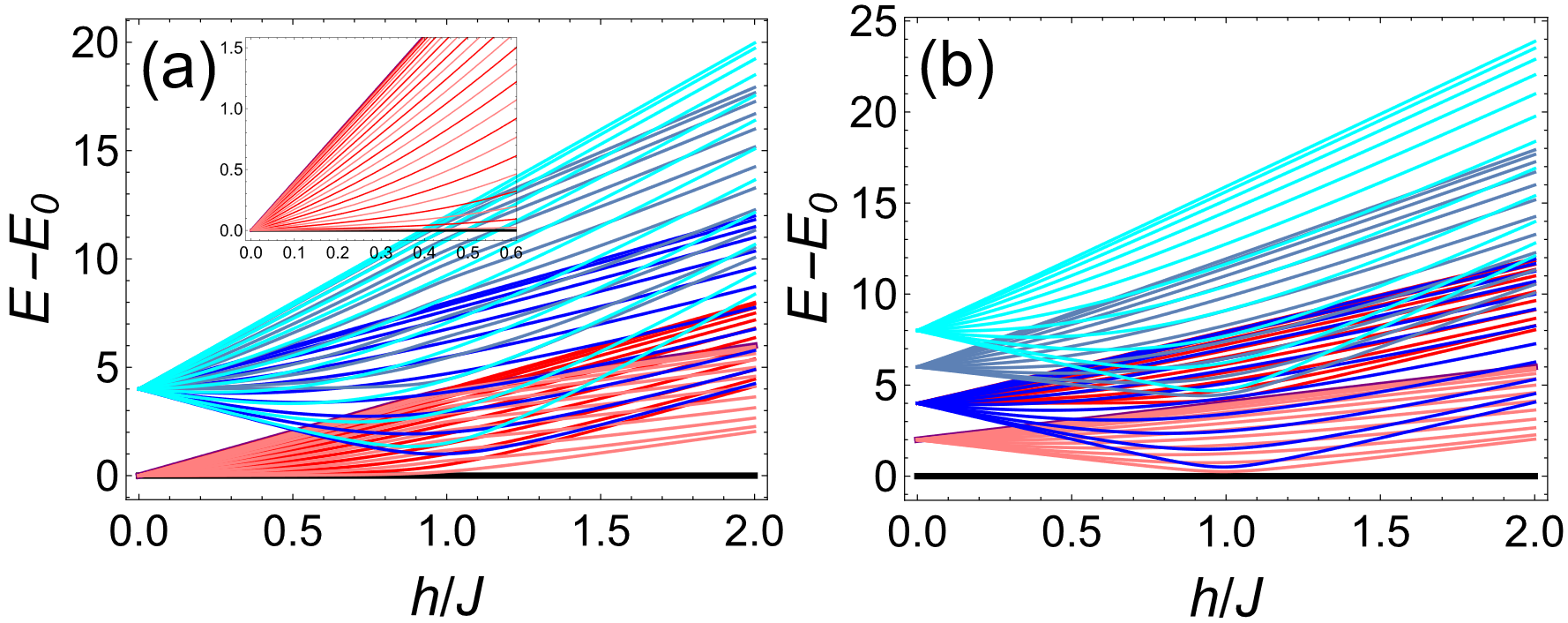}
  \caption{The low energy of the generalized quantum cluster model with quantum fluctuations with (a) $m=3$ and $N=25$ ; (b) $m=4$ and $N = 25$. $E_{0}$ represents the energy of ground state.}
  \label{figspectrum}
\end{figure}
In this section, we focus on the generalized quantum cluster model with large quantum fluctuations.

With the inclusion of a large transverse field, we find that for situations of $N$ and $m$,
the critical point occurs at $ h = J $, where the system undergoes a quantum phase transition.
However, in the special case of $ m \in \text{odd} $ and $ N \in \text{odd} $,
the low-energy spectrum remains gapless for $ h < J $.
This gapless feature originates from the $2N$-fold degenerate ground states, whose degeneracy is partially lifted by the transverse field,
as illustrated in Fig.~\ref{figspectrum}(a).
In contrast, for all other parameter combinations, the system remains gapped in the $ h < J $ regime,
as shown in Fig.~\ref{figspectrum}(b).
When the transverse field becomes dominant ($ h > J $), the system enters the paramagnetic phase,
and an excitation gap opens again.

\begin{figure}[t]
  \centering
  \includegraphics[width=\columnwidth]{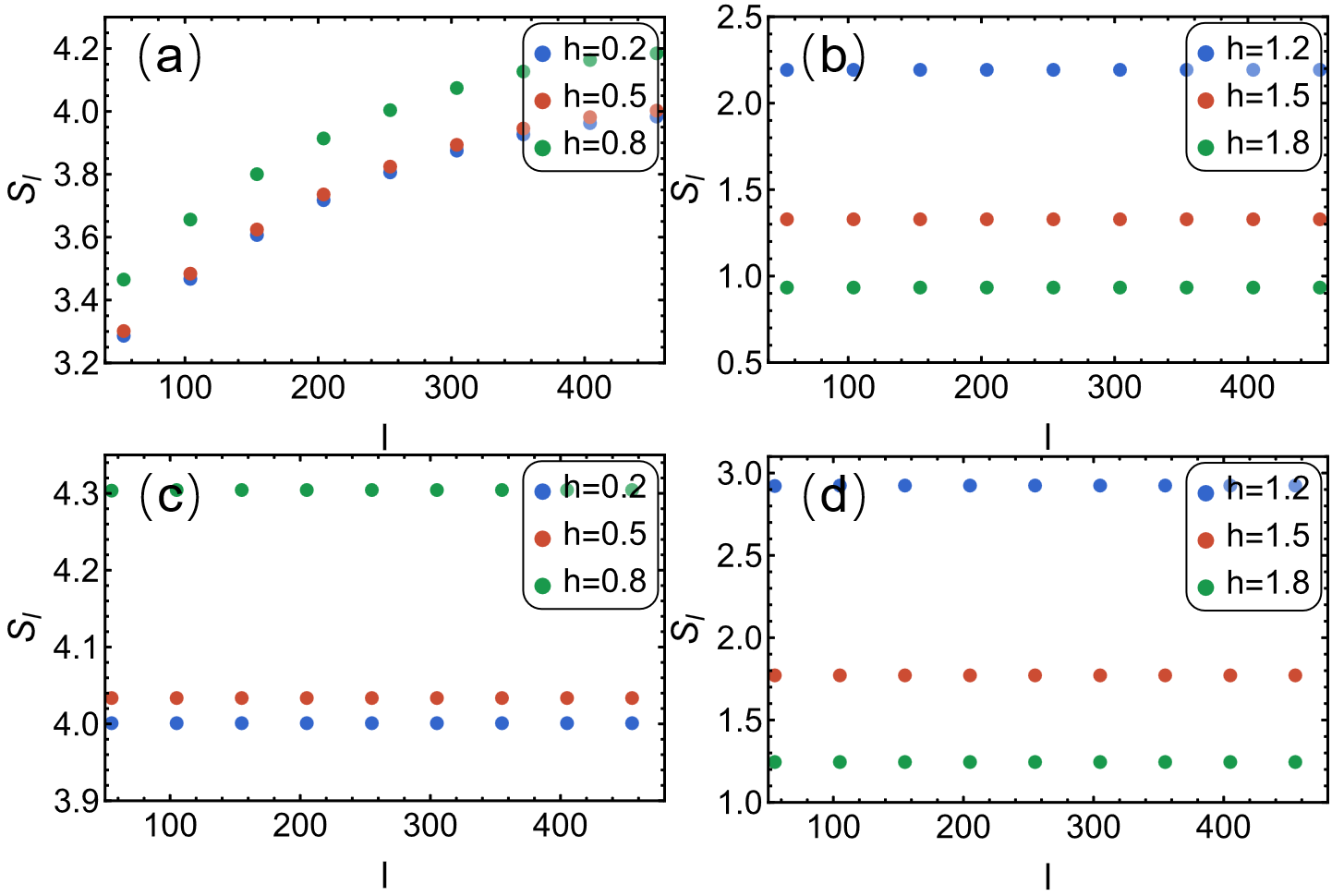}
  \caption{Entanglement entropy $S_l$ as a function of the block size $l$ $(l\geq m)$ with (a)-(b) $N=1001$ and $m=3$, and (c)-(d) $N=1001$ and $m=4$.}
  \label{figEEh}
\end{figure}

\begin{figure}[t]
  \centering
  \includegraphics[width=\columnwidth]{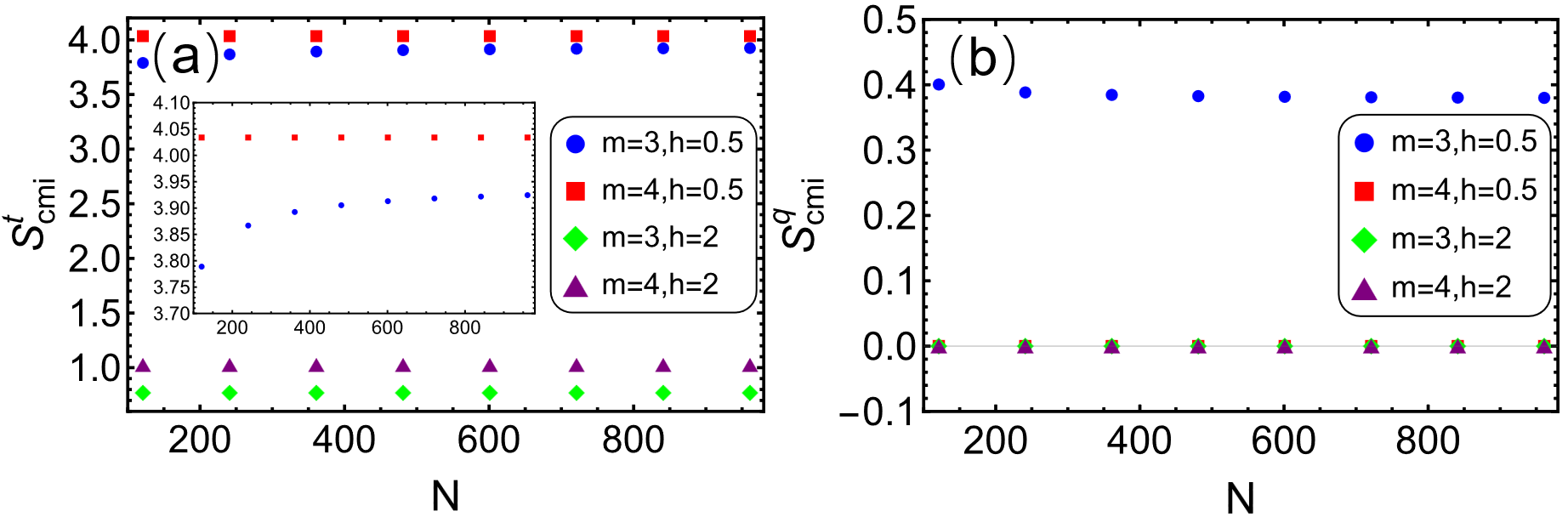}
  \caption{Quantum mutual information entropy as a function of system size for the two representative cases.}
  \label{figCMIh}
\end{figure}
Next, we mainly focus on how quantum fluctuations affect the entanglement entropy, as well as the three- and four-part conditional mutual information entropy. For convenience, we set $J=1$ as an unite.

In Fig.\ref{figEEh}, we show the behavior of the entanglement entropy as a function of the block size under different strengths of the transverse field.
We find that when both the system size $N$ and the interaction range $m$ are odd, the entanglement entropy exhibits a distinct non-local character for $h < J$.
It increases monotonically with the block size and reaches its maximum when the block covers half of the total system size, as shown in Fig.\ref{figEEh}(a).
In contrast, for all other combinations of
 $N$ and $m$, the entanglement entropy remains nearly constant with respect to the block size in the $h < J$ regime as show in Fig.\ref{figEEh}(c), indicating the absence of non-local entanglement.

When the transverse field becomes strong ($h > J$), the entanglement entropy no longer depends on the block size for any parameter set
and saturates to a constant value, corresponding to a paramagnetic phase with the absence of non-local entanglement, as show in Fig.\ref{figEEh}(b) and Fig.\ref{figEEh}(d).
Therefore, only in the $N, m \in \text{odd}$ case and for $h < J$ does the system maintain non-local entanglement characteristics.

Furthermore, the phase transition can also be characterized by the behavior of the entanglement entropy.
At the critical point $h = J$, the entanglement entropy exhibits a divergent scaling behavior consistent with
the prediction of conformal field theory (CFT) in the Ising universality class.
Specifically, the scaling follows the form $\frac{m}{6}\log_{2}(N)$, where the coefficient corresponds to the
central charge of the underlying critical theory.
The detailed numerical verification and fitting procedure are provided in the Appendix \ref{appendixCC}.

Next, we investigate how quantum fluctuations influence the three-part and four-part conditional mutual information entropies.
As shown in Figs.\ref{figCMIh}(a) and \ref{figCMIh}(b), the tripartite conditional mutual information entropy $S_{cmi}^{t}$
remains finite in both representative parameter cases when a moderate transverse field is introduced.
However, when both the interaction range $m$ and the system size $N$ are odd,
the three-part condition mutual information entropy $S_{cmi}^{t}$ also increases with the system size,
indicating the growth of non-local properties.
In contrast, for other cases, $S_{cmi}^{t}$ saturates to a constant value as the system size increases,
consistent with the one-dimensional area law of entanglement.

In contrast, a qualitatively different feature emerges in the four-part conditional mutual information entropy.
As shown in Fig.~\ref{figCMIh}(a), when both the interaction range $m$ and the system size $N$ are odd,
the four-part entropy $S_{cmi}^{q}$ remains finite for $h < J$.
This non-vanishing behavior demonstrates the persistence of non-local entanglement in the presence of quantum fluctuations.
Because $S_{cmi}^{q}$ effectively cancels local contributions, its finite value directly reflects genuine long-range quantum correlations and entanglement.
In all other parameter regimes, as shown in Fig.\ref{figCMIh}(a), $S_{cmi}^{q}$ is equal to zero, confirming the absence of long-range entanglement in those cases.

In Fig.\ref{figCMIh}, we present the behavior of the tripartite and quadripartite conditional mutual information in the paramagnetic phase. For the regime $ h > J $, we find that the tripartite conditional mutual information entropy
remains constant for all combinations of $ m $ and $ N $,
while the four-part mutual information entropy vanishes.
As the transverse field becomes much stronger $(h \gg J)$,
the tripartite entropy also approaches zero,
indicating that the ground state in this regime is a disentangled paramagnetic phase.

The above results demonstrate that quantum fluctuations modify, but do not completely destroy,
the system’s entanglement structure.
The tripartite mutual information reflects stable short-range correlations,
whereas the four-part entropy selectively captures long-range entanglement.
In particular, for $m, N \in \text{odd}$, the survival of a finite $S_{cmi}^{q}$ under $h < J$ confirms that
the system retains its non-local entanglement and topological character even in the presence of a transverse field.
As a special case, when $m = 1$, the model reduces to the transverse-field Ising chain,
for which the persistence of long-range entanglement in the phase has been explicitly demonstrated in Ref.\cite{Odavic_2023_Quantum,Torre_2024_scicore}.

\section{Summary and discussion}\label{sec:conclusion}
In this work, we have investigated the entanglement structure and quantum phase transitions in a generalized quantum cluster model.
By analyzing the entanglement entropy and conditional mutual information under different subsystem partitions, we identified distinct signatures of long-range entanglement that depend crucially on the interaction range $m$ and the system size $N$.
When both $m$ and $N$ are odd, the system exhibits nontrivial entanglement features, manifested by long-range correlation and genuine long-range entanglement in infinitesimal but finite field. Even in the presence of a large transverse field, the system with odd $m$ and odd
$N$ retains nonvanishing multipartite conditional mutual information, reflecting the persistence of long-range entanglement.
In contrast, all other parameter combinations yield fully gapped phases devoid of long-range correlations.
These results show that the system size and the interaction range play a decisive role in determining both the emergence and the stability of long-range quantum entanglement in one-dimensional cluster systems.

To elucidate the physical origin of the highly degenerate ground state ($2N$-fold) and the emergent long-range entanglement when both $N$
and m are odd, we consider the limiting case of $m=1$. In this limit, the model reduces to the one-dimensional nearest-neighbor antiferromagnetic Ising model under periodic boundary conditions. Due to the odd system size, the spins are unable to satisfy the antiferromagnetic alignment globally, subjecting the system to geometric frustration. This constraint directly results in a large ground-state degeneracy and long-range entanglement in infinitesimal but finite field \cite{Torre_2024_scicore}.  Extending this picture, one can intuitively envision that for odd m ($m>1$),  model gives rise to a similar form of “frustration.”  and develops a highly degenerate ground-state manifold accompanied by the emergence of long-range entanglement in a finite transverse field.

\section*{Acknowledgments}
We thank Duan-Lu Zhou for useful discussions. This work is supported by National Key Research and Development Program of China (Grant No.2021YFA1402104)
and the NSFC under Grants  No.12474287 and No.12547107 and No.T2121001.

\section*{DATA AVAILABILITY}
The data that support the findings of this
article are openly available \cite{data}.

\appendix

\section{Probing ground-state degeneracy through Kramers-Wannier transformation}\label{appendixKWT}
To systematically investigate ground-state degeneracy in generalized quantum cluster models, we employ the Kramers-Wannier (KW) transformation, a non-local mapping that preserves the Pauli algebra while revealing hidden structural symmetries. Crucially, the parity of the interaction range $m$ dictates fundamentally different Hamiltonian structures:

For even $m$, the duality transformation $Z_j = \sigma_j^x \tau_{m,j}^z \sigma_{j+m}^x$, $X_j = \sigma_j^x$ yields the simplified Hamiltonian
$
H = J \sum_{j=1}^N Z_j.
$
This represents a direct sum of mutually commuting operators whose spectrum is trivially solvable. The absence of non-commuting terms ensures a unique ground state configuration for all system sizes $N$, with energy minimization requiring simultaneous eigenvalue minimization of each $Z_j$ operator.

Conversely, odd $m$ introduces topological complications manifested through boundary effects. The transformation now requires distinct boundary operators: $Z_j = \sigma_j^x \tau_{m,j}^z \sigma_{j+m}^x$ for $1 \leq j \leq N-1$, while the terminal operator becomes $Z_N = \sigma_N^x \prod_{l=1}^{m-1} \sigma_l^z$. This asymmetric mapping produces the constrained Hamiltonian:
\begin{equation}
H = J \sum_{j=1}^{N-1} Z_j + J \prod_{j=1}^{N-1} Z_j.
\end{equation}
The non-local product term $\prod_{j=1}^{N-1} Z_j$ imposes a global constraint that effectively reduces the number of independent $Z_j$ operators to $N-1$. This constraint gives rise to a macroscopic ground-state degeneracy with a striking dependence on the system size. For even $N$, the degeneracy is two-fold, consistent with spontaneous $\mathbb{Z}_2$ symmetry breaking. In contrast, for odd $N$, the degeneracy increases to $2N$.



\section{Non-locality in entanglement entropy}\label{appendixEE}
The entanglement entropy is defined via the reduced density matrix as
\begin{equation}
S_l = - \mathrm{tr} \left( \rho_l \log_2 \rho_l \right),
\end{equation}
where $\rho_l$ is the reduced density matrix with a block of $l$ spins\cite{Vidal_2003_PRL} and $ l \geq m$ is chosen for computational convenience and to guarantee that the entanglement entropy reaches its saturation value. Following the method of Ref.\cite{Vidal_2003_PRL,
Latorre_2003_arxiv}, $S_l$ can be computed from the correlation matrix $\Gamma_l$ as
\begin{equation}
S_l = - \sum_{j=1}^{2l} f(v_j) = - \sum_{j=1}^{2l} \frac{1+v_j}{2} \log_2 \frac{1+v_j}{2},
\label{eq:entropy}
\end{equation}
where $v_j$ are the imaginary parts of the eigenvalues of $\Gamma_l$, which takes the block-Toeplitz form
\begin{equation}
\Gamma_{l} =
\begin{vmatrix}
\Pi_{0} & \Pi_{1} & \cdots & \Pi_{l-1} \\
\Pi_{-1} & \Pi_{0} & \cdots & \Pi_{l-2} \\
\vdots & \vdots & \ddots & \vdots \\
\Pi_{1-l} & \Pi_{2-l} & \cdots & \Pi_{0}
\end{vmatrix},
\end{equation}
with $2\times 2$ block elements
\begin{equation}
\Pi_{r} =
\begin{vmatrix}
0 & -g_{r} \\
g_{-r} & 0
\end{vmatrix}.
\end{equation}
The $g_{r}$ represents the correlation which is defined as $g_{r}=\langle (c_{j}^{\dag}-c_{j})(c_{j+r}^{\dag}+c_{j+r}) \rangle$.

Firstly, we consider the case with $N\in \text{odd}$ and $m\in \text{odd}$ in infinitesimal but finite field.
Without loss of generality, we focus on the definite-parity and  momentum ground state $c_0^\dag|\varphi^+\rangle$, which satisfies the translational symmetry, within the $2N$-fold degenerate ground-state manifold. Remarkably, all other degenerate ground states which is definite-parity and momentum, for example the state $|\varphi^-\rangle$, exhibit identical entanglement properties and numerical validation is shown in Appendix.\ref{appendixss}. Therefore, the specific choice of ground state does not affect our conclusions.

The elements of the correlation matrix $ \Gamma_l $ take the form,
\begin{equation}
g_r = (-1)^m \delta_{-r, m} + \frac{2}{N}, \label{eq:gr}
\end{equation}
Although the term $ \frac{2}{N} $ vanishes in the strict thermodynamic limit, it cannot be neglected here. This is because the size of the correlation matrix scales linearly with $ N $, and hence the cumulative contribution of this term remains finite.

Due to the effect of the term  $\frac{2}{N} $, the eigenvalues of the correlation matrix are given by
\begin{equation}
\left\{
\underbrace{0,\dots,0}_{2(m-1)}, \underbrace{-1,1,\dots,-1,1}_{2(l-m-1)}, v_1, -v_1, v_2, -v_2
\right\}, \label{eq:eigenvalues}
\end{equation}
where
\begin{equation}
\begin{gathered}
v_{1} = \frac{\sqrt{2}}{2} \sqrt{4 (\alpha^{2} + \beta - \alpha)+ 1 +(1- 2 \alpha) x},  \\
v_{2} = \frac{\sqrt{2}}{2} \sqrt{4 (\alpha^{2} + \beta - \alpha)+ 1 + (2 \alpha- 1) x},  \\
x = \sqrt{4 (\alpha^2 + 2 \beta - \alpha) + 1},
\end{gathered}
\end{equation}
where $\alpha=\frac{l}{N}, \beta=\frac{m}{N}$. Notice that the factor $\alpha$ will propose the non-local information in entanglement. Then, the entanglement entropy is given by,
\begin{equation}
S_{l} = m-1 - f(v_1) - f(-v_1) - f(v_2) - f(-v_2). \label{eq:entropyS}
\end{equation}
In the thermodynamic limit ($N\rightarrow\infty$), the entanglement entropy is satisfied:
\begin{equation}
m \leq \lim_{N\rightarrow\infty} S_{l} \leq m+1 \quad \text{for}
\quad m\leq l\leq \frac{N-1}{2}.\label{eq:entropySinf}
\end{equation}

In comparison, we also analyze the rest of the cases of the model. The ground state is chosen by $|\varphi^{-}\rangle$ for convenience.
Specifically, under these conditions, the elements of correlation matrix are the same and can be simplified as
\begin{equation}
g_r = (-1)^m \delta_{-r, m},
\end{equation}
and the eigenvalues of the correlation matrix are given by
\begin{equation}
\left\{
\underbrace{0,\dots,0}_{2m}, \underbrace{-1,1,\dots,-1,1}_{2(l-m)},
\right\}, \label{eq:eigenvalues}
\end{equation}

The entanglement entropy can be computed exactly, yielding
\begin{equation}
S_{l} = -m \log_2 \left( \frac{1}{2} \right) = m. \label{eq:entropyT}
\end{equation}

\begin{figure}[t]
  \centering
  \includegraphics[width=0.9\columnwidth]{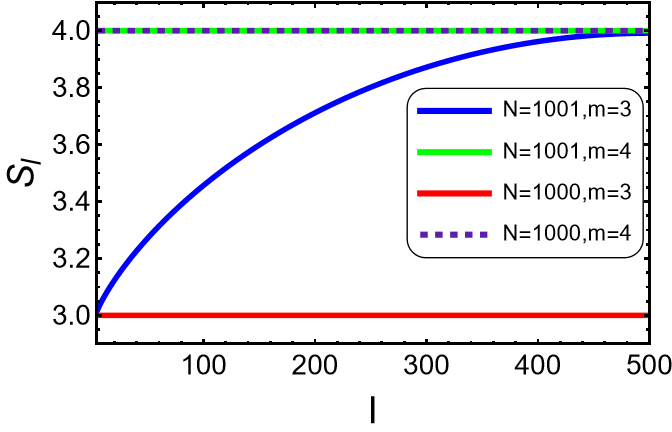}
  \caption{ Entanglement entropy $S_l$ as a function of the block size $l$ $(l\geq m)$.}
  \label{figEE}
\end{figure}

From Eq.~(\ref{eq:entropySinf}) and Eq.~(\ref{eq:entropyT}), we observe that although the entanglement entropy in both cases satisfies an area law in the thermodynamic limit, there remains a significant difference in entanglement properties. Specifically, when both $N$ and $m$ are odd, the entanglement entropy depends not only on the interaction range $m$ itself, but also on the non-local factor $\alpha $, i.e., the fraction between the block size $l$ and the total system size $N$. As shown in Fig.\ref{figEE}, we consider $N=1001$ and $m=3$ as an example and the entropy (blue line) increases with the block size $l$, for $l = m$, it takes the value $m$, and when the block size $l$ approaches half of the system size, it reaches its maximum $m+1$.
Hence, the entanglement entropy does not saturate but continues to grow with the subsystem size. Consequently, the entanglement is not solely determined by boundary degrees of freedom but is also tied to global features, highlighting the system's long-range correlations.

In contrast, for all other cases, the entanglement entropy depends solely on the interaction range
$m$ and remains independent of the block size
$l$, exhibiting a clear saturation plateau in full agreement with Eq.~(\ref{eq:entropyT}), as shown in Fig.~\ref{figEE}, which indicates that the entanglement is confined to the boundary degrees of freedom and the presence of short-range correlations.

Here we summarize the main properties of the entanglement entropy. Although in both cases the entanglement entropy obeys an area-law scaling, we find that when $m$ is odd and the system size $N$ is also odd, the entanglement entropy exhibits a nontrivial dependence on the subsystem size. This behavior indicates that the entanglement contribution in this regime may not originate solely from the boundary between subsystems, but may also involve correlations within the bulk of the subsystem. In other words, the system displays nonlocal features and suggests the presence of long-range correlations. However, at this stage, we cannot unambiguously conclude that the system hosts long-range entanglement, since the contribution from boundary  has not yet been completely disentangled.

\section{Same behaviour of entanglement entropy for ground-states in infinitesimal but finite field}\label{appendixss}
\begin{figure}[h]
  \centering
  \includegraphics[width=0.9\columnwidth]{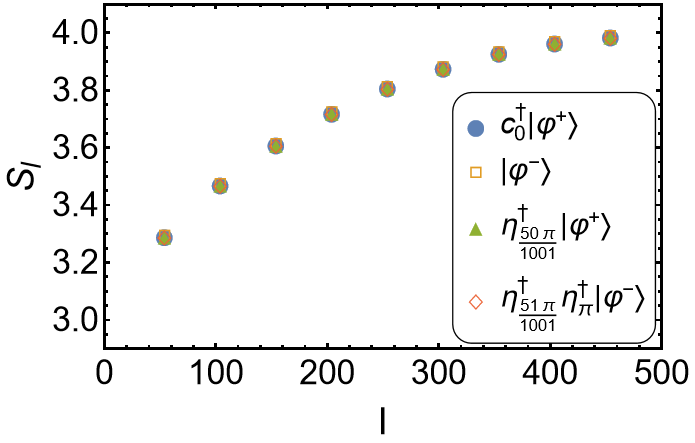}
  \caption{ Entanglement entropy $S_l$ as a function of the block size $l$ $(l\geq m)$ for different states with $N=1001,m=3$.}
  \label{figstate}
\end{figure}

In this section, we numerically evaluate the entanglement entropy to verify that all $2N$-fold degenerate ground states with well-defined momentum exhibit identical entanglement behavior. As shown in Fig.\ref{figstate}, we set $N=1001,m=3$ and explicitly consider four representative ground states with different momenta, including two with odd parity $c_{0}^{\dag}|\varphi^{+}\rangle; \eta_{\frac{50\pi}{1001}}^{\dag}|\varphi^{+}\rangle$ and two with even parity $|\varphi^{-}\rangle; \eta_{\frac{51\pi}{1001}}^{\dag}\eta_{\pi}^{\dag}|\varphi^{-}\rangle$. Despite their distinct fermion excition, the corresponding entanglement entropies coincide exactly.

This result demonstrates that the entanglement properties are identical for all $2N$-fold degenerate ground states with defined momentum and parity.

\section{Details of three-part conditional mutual information}\label{appendixTPMIC}

\begin{figure}[t]
  \centering
  \includegraphics[width=0.85\columnwidth]{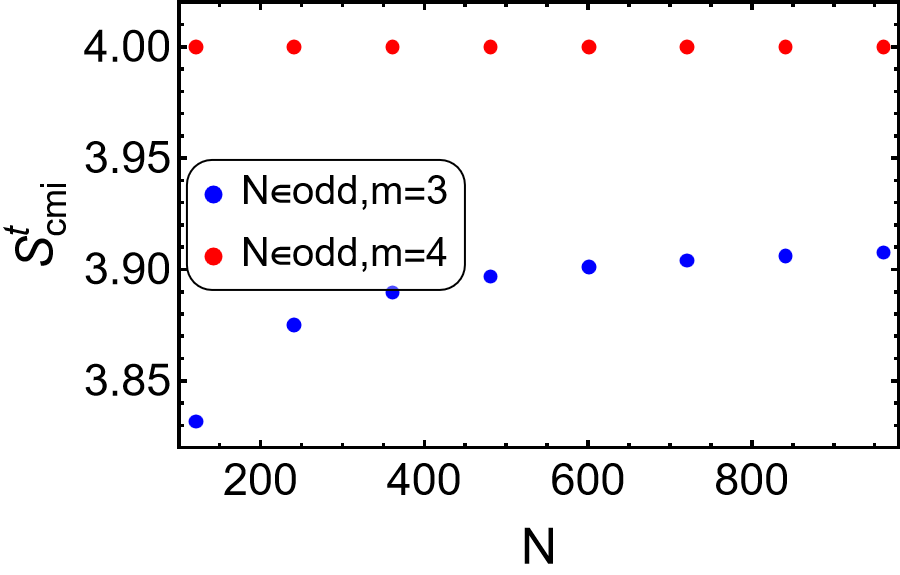}
  \caption{Three-part quantum mutual information entropy as a function of system size.}
  \label{figCMIt}
\end{figure}

When the system is partitioned into three subsystems, the conditional quantum mutual information entropy captures only partial information. Although $S_{cmi}^{t}$ remains nonzero in both case as shown in Fig.\ref{figCMIt}, it approaches a constant value in the thermodynamic limit, consistent with the 1D area law of entanglement.
However, we find that for the case where both the interaction range $m$ and the system size $N$ are odd, the three-part conditional mutual information entropy increases with system size under the
approximately equal partitioning scheme. In contrast, for all other parameter combinations,
$S_{cmi}^{t}$ remains constant regardless of system size.
This size-dependent growth of $S_{cmi}^{t}$ also reflects the presence of long-range correlation. However, similarly to the entanglement entropy, the three-part mutual information still contains local contributions. Therefore, at this stage, we cannot unambiguously conclude that the system hosts long-range entanglement.

\begin{table}[htbp]
\centering
\small
\begin{tabular}{@{}p{0.3\columnwidth}p{0.2\columnwidth}p{0.4\columnwidth}@{}}
\toprule
\textbf{Situation} & \textbf{$S_{l}$} & \textbf{$S_{cmi}$} \\
\midrule
$m \in \mathrm{odd}, N \in \mathrm{odd}$ & Non-local & $S_{\mathrm{cmi}}^{t} \neq 0, \; S_{\mathrm{cmi}}^{q} \neq 0$ \\
Other cases & local & $S_{\mathrm{cmi}}^{t} \neq 0, \; S_{\mathrm{cmi}}^{q} = 0$ \\
\bottomrule
\end{tabular}
\caption{Entanglement entropy and quantum condition mutual information across scenarios}
\label{tab:quantum_mutual_information}
\end{table}
\normalsize

Table~\ref{tab:quantum_mutual_information} summarizes the results in infinitesimal but finite field. When both $m$ and $N$ are odd, the system exhibits a distinctly non-local entanglement structure, as verified by the entanglement entropy and quantum mutual information. In this case, the four-part quantum mutual information remains finite, which clearly indicates the presence of long-range quantum correlations. In contrast, for all other cases, including even $m$ with odd $N$, the four-part quantum mutual information vanishes identically. This vanishing behavior reflects a short-range entanglement structure, confirming the absence of long-range entanglement in these regimes.

\section{Scaling behaviour of entanglement entropy at critical point}\label{appendixCC}
This appendix aims to provide the theoretical foundation for the analysis of the relationship between entanglement entropy and the total system size presented in the main text. Specifically, we compute the entanglement entropy when the system is bi-partitioned into two equal halves. Since we choose the total system size $N$ to be odd, the subsystem size is taken as
$l=(N-1)/2$. We then investigate how the entanglement entropy evolves as the total system size increases.
\begin{figure}[t]
\centering
\includegraphics[width=\columnwidth]{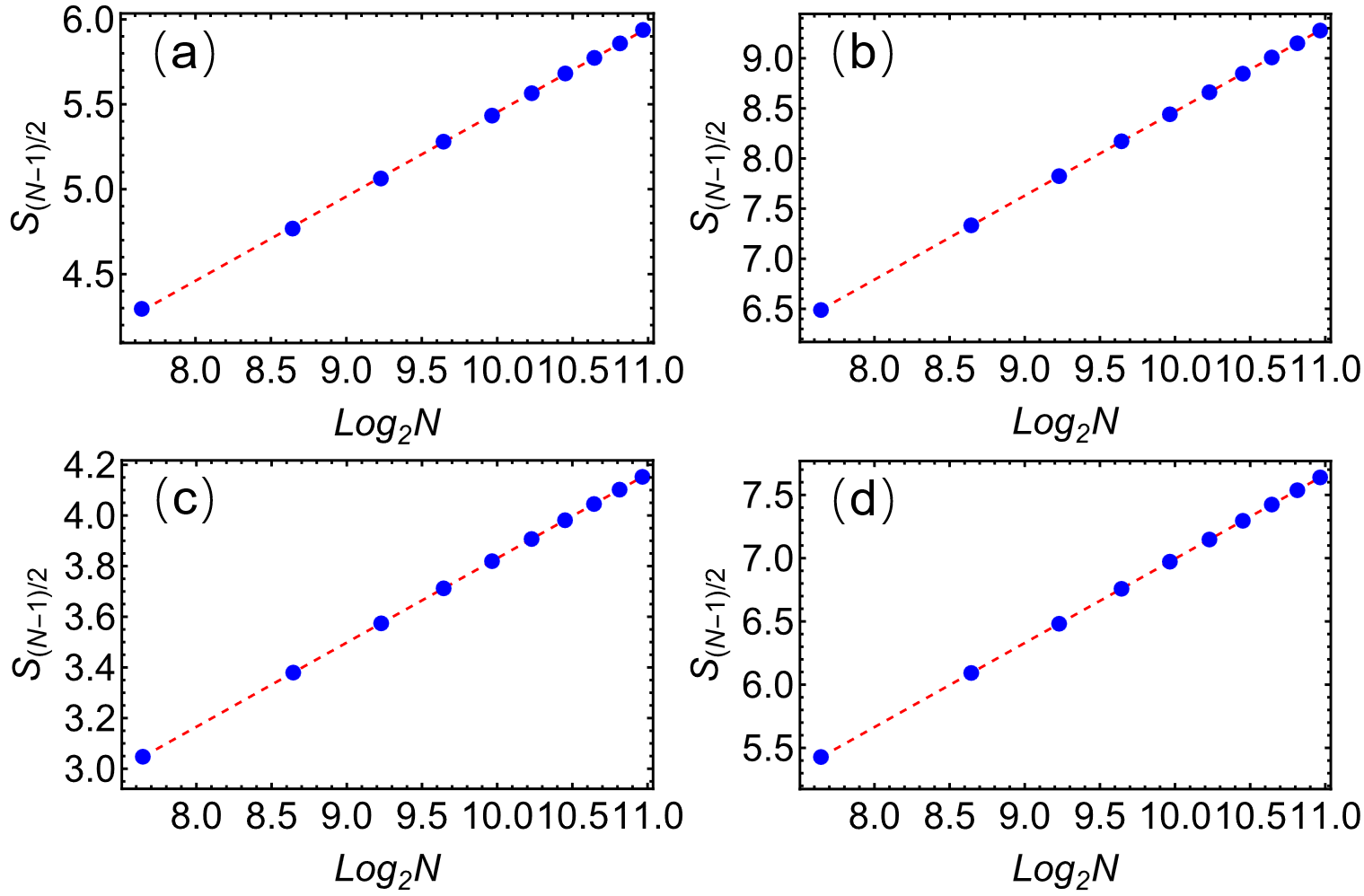}
\caption{Scaling behavior of the entanglement entropy at the critical point
$h=J$. Panels (a)–(d) display the convergence of the scaling behavior for interaction ranges $m=3$, $m=5$, $m=2$, and $m=4$, respectively. Blue dots represent the numerical data points, while the red dashed lines correspond to the fitting curves.}
\label{figEEscaling}
\end{figure}

In Fig.~\ref{figEEscaling}, we illustrate the scaling behavior of the entanglement entropy at the quantum critical point $h=J$. By performing finite-size scaling analysis of the entanglement entropy, we find that it can be well fitted by the following logarithmic forms:
$S_{(N-1)/2}=0.49655\times \log_{2}{(N)}+0.4881$ for Fig.~\ref{figEEscaling}(a),
$S_{(N-1)/2}=0.8389\times \log_{2}{(N)}+0.0798$ for Fig.~\ref{figEEscaling}(b),
$S_{(N-1)/2}=0.3327\times \log_{2}{(N)}+0.5035$ for Fig.~\ref{figEEscaling}(c), and
$S_{(N-1)/2}=0.6667\times \log_{2}{(N)}+0.3404$ for Fig.~\ref{figEEscaling}(d).
These results demonstrate that the entanglement entropy at the critical point exhibits a logarithmic divergence with system size, following the scaling relation
$S_{(N-1)/2} \sim \frac{m}{6} \log_{2}(N)$, where $m$ denotes the interaction range of the generalized cluster model.

This logarithmic scaling is in excellent agreement with the prediction of conformal field theory (CFT) for systems in the Ising universality class. Specifically, the coefficient of the logarithm is directly related to the central charge of the underlying critical theory, reflecting the universal nature of the quantum critical point. The results further indicate that longer interaction ranges $m$ lead to proportionally larger entanglement growth, consistent with the intuitive picture that the extent of correlations in the system increases with the interaction range. Overall, this analysis confirms that the generalized cluster model exhibits critical behavior fully captured by CFT predictions, with the entanglement entropy serving as a sensitive probe of universal properties at the phase transition.

\bibliography{GCM_bib}

@PREAMBLE{
 "\providecommand{\noopsort}[1]{}" 
 # "\providecommand{\singleletter}[1]{#1}%" 
}

@article{Vidal_2003_PRL,
  title = {Entanglement in Quantum Critical Phenomena},
  author = {Vidal, G. and Latorre, J. I. and Rico, E. and Kitaev, A.},
  journal = {Phys. Rev. Lett.},
  volume = {90},
  issue = {22},
  pages = {227902},
  numpages = {4},
  year = {2003},
  month = {Jun},
  publisher = {American Physical Society},
  doi = {10.1103/PhysRevLett.90.227902},
  url = {https://link.aps.org/doi/10.1103/PhysRevLett.90.227902}
}

@article{Latorre_2003_arxiv,
  title = {Ground state entanglement in quantum spin chains},
  author = {Latorre, J. I. and Rico, E. and Vidal, G.},
  journal = {Quant. Inf. Comput.},
  volume = {4},
  number = {1},
  pages = {48--92},
  year = {2004}
}

@book{Fradkin_2013_field,
  title={Field Theories of Condensed Matter Systems},
  author={E. Fradkin},
  year={2013},
  publisher={Cambridge University Press, Cambridge}
}

@article{Calabrese_2009_JPA,
doi = {10.1088/1751-8113/42/50/504005},
url = {https://doi.org/10.1088/1751-8113/42/50/504005},
year = {2009},
month = {dec},
publisher = {},
volume = {42},
number = {50},
pages = {504005},
author = {Calabrese, Pasquale and Cardy, John},
title = {Entanglement entropy and conformal field theory},
journal = {Journal of Physics A: Mathematical and Theoretical}
}

@article{Eisert_2010_RMP,
  title = {Colloquium: Area laws for the entanglement entropy},
  author = {Eisert, J. and Cramer, M. and Plenio, M. B.},
  journal = {Rev. Mod. Phys.},
  volume = {82},
  issue = {1},
  pages = {277--306},
  numpages = {0},
  year = {2010},
  month = {Feb},
  publisher = {American Physical Society},
  doi = {10.1103/RevModPhys.82.277},
  url = {https://link.aps.org/doi/10.1103/RevModPhys.82.277}
}

@article{Amico_2008_RMP,
  title = {Entanglement in many-body systems},
  author = {Amico, Luigi and Fazio, Rosario and Osterloh, Andreas and Vedral, Vlatko},
  journal = {Rev. Mod. Phys.},
  volume = {80},
  issue = {2},
  pages = {517--576},
  numpages = {0},
  year = {2008},
  month = {May},
  publisher = {American Physical Society},
  doi = {10.1103/RevModPhys.80.517},
  url = {https://link.aps.org/doi/10.1103/RevModPhys.80.517}
}

@article{HAMMA_2005_PLA,
title = {Ground state entanglement and geometric entropy in the Kitaev model},
journal = {Physics Letters A},
volume = {337},
number = {1},
pages = {22-28},
year = {2005},
issn = {0375-9601},
doi = {https://doi.org/10.1016/j.physleta.2005.01.060},
url = {https://www.sciencedirect.com/science/article/pii/S0375960105001544},
author = {Alioscia Hamma and Radu Ionicioiu and Paolo Zanardi}
}

@article{Levin_2006_PRL,
  title = {Detecting Topological Order in a Ground State Wave Function},
  author = {Levin, Michael and Wen, Xiao-Gang},
  journal = {Phys. Rev. Lett.},
  volume = {96},
  issue = {11},
  pages = {110405},
  numpages = {4},
  year = {2006},
  month = {Mar},
  publisher = {American Physical Society},
  doi = {10.1103/PhysRevLett.96.110405},
  url = {https://link.aps.org/doi/10.1103/PhysRevLett.96.110405}
}

@article{Kitaev_2006_PRL,
  title = {Topological Entanglement Entropy},
  author = {Kitaev, Alexei and Preskill, John},
  journal = {Phys. Rev. Lett.},
  volume = {96},
  issue = {11},
  pages = {110404},
  numpages = {4},
  year = {2006},
  month = {Mar},
  publisher = {American Physical Society},
  doi = {10.1103/PhysRevLett.96.110404},
  url = {https://link.aps.org/doi/10.1103/PhysRevLett.96.110404}
}

@article{Wen_2017_RMP,
  title = {Colloquium: Zoo of quantum-topological phases of matter},
  author = {Wen, Xiao-Gang},
  journal = {Rev. Mod. Phys.},
  volume = {89},
  issue = {4},
  pages = {041004},
  numpages = {17},
  year = {2017},
  month = {Dec},
  publisher = {American Physical Society},
  doi = {10.1103/RevModPhys.89.041004},
  url = {https://link.aps.org/doi/10.1103/RevModPhys.89.041004}
}

@article{
Wen_2019_Sci,
author = {Xiao-Gang Wen},
title = {Choreographed entanglement dances: Topological states of quantum matter},
journal = {Science},
volume = {363},
number = {6429},
pages = {eaal3099},
year = {2019},
doi = {10.1126/science.aal3099},
URL = {https://www.science.org/doi/abs/10.1126/science.aal3099}
}

@article{Alioscia_2005_PRA,
  title = {Bipartite entanglement and entropic boundary law in lattice spin systems},
  author = {Hamma, Alioscia and Ionicioiu, Radu and Zanardi, Paolo},
  journal = {Phys. Rev. A},
  volume = {71},
  issue = {2},
  pages = {022315},
  numpages = {10},
  year = {2005},
  month = {Feb},
  publisher = {American Physical Society},
  doi = {10.1103/PhysRevA.71.022315},
  url = {https://link.aps.org/doi/10.1103/PhysRevA.71.022315}
}

@article{Li_2008_PRL,
  title = {Entanglement Spectrum as a Generalization of Entanglement Entropy: Identification of Topological Order in Non-Abelian Fractional Quantum Hall Effect States},
  author = {Li, Hui and Haldane, F. D. M.},
  journal = {Phys. Rev. Lett.},
  volume = {101},
  issue = {1},
  pages = {010504},
  numpages = {4},
  year = {2008},
  month = {Jul},
  publisher = {American Physical Society},
  doi = {10.1103/PhysRevLett.101.010504},
  url = {https://link.aps.org/doi/10.1103/PhysRevLett.101.010504}
}

@article{Pollmann_2010_PRB,
  title = {Entanglement spectrum of a topological phase in one dimension},
  author = {Pollmann, Frank and Turner, Ari M. and Berg, Erez and Oshikawa, Masaki},
  journal = {Phys. Rev. B},
  volume = {81},
  issue = {6},
  pages = {064439},
  numpages = {10},
  year = {2010},
  month = {Feb},
  publisher = {American Physical Society},
  doi = {10.1103/PhysRevB.81.064439},
  url = {https://link.aps.org/doi/10.1103/PhysRevB.81.064439}
}

@article{Fidkowski_2010_PRL,
  title = {Entanglement Spectrum of Topological Insulators and Superconductors},
  author = {Fidkowski, Lukasz},
  journal = {Phys. Rev. Lett.},
  volume = {104},
  issue = {13},
  pages = {130502},
  numpages = {4},
  year = {2010},
  month = {Apr},
  publisher = {American Physical Society},
  doi = {10.1103/PhysRevLett.104.130502},
  url = {https://link.aps.org/doi/10.1103/PhysRevLett.104.130502}
}

@article{Turner_2011_PRB,
  title = {Topological phases of one-dimensional fermions: An entanglement point of view},
  author = {Turner, Ari M. and Pollmann, Frank and Berg, Erez},
  journal = {Phys. Rev. B},
  volume = {83},
  issue = {7},
  pages = {075102},
  numpages = {11},
  year = {2011},
  month = {Feb},
  publisher = {American Physical Society},
  doi = {10.1103/PhysRevB.83.075102},
  url = {https://link.aps.org/doi/10.1103/PhysRevB.83.075102}
}

@book{Zeng_2019_quantum,
  title = {Quantum Information Meets Quantum Matter},
  author = {Zeng, Bei and Chen, Xie and Zhou, Duan-Lu and Wen, Xiao-Gang},
  year = {2019},
  publisher = {Springer},
  address = {New York}
}

@article{Fromholz_2020_PRB,
  title = {Entanglement topological invariants for one-dimensional topological superconductors},
  author = {Fromholz, P. and Magnifico, G. and Vitale, V. and Mendes-Santos, T. and Dalmonte, M.},
  journal = {Phys. Rev. B},
  volume = {101},
  issue = {8},
  pages = {085136},
  numpages = {12},
  year = {2020},
  month = {Feb},
  publisher = {American Physical Society},
  doi = {10.1103/PhysRevB.101.085136},
  url = {https://link.aps.org/doi/10.1103/PhysRevB.101.085136}
}

@Article{Micallo_2020_scicore,
	title={{Topological entanglement properties of disconnected partitions in the Su-Schrieffer-Heeger model}},
	author={Tommaso Micallo and Vittorio Vitale and Marcello Dalmonte and Pierre Fromholz},
	journal={SciPost Phys. Core},
	volume={3},
	pages={012},
	year={2020},
	publisher={SciPost},
	doi={10.21468/SciPostPhysCore.3.2.012},
	url={https://scipost.org/10.21468/SciPostPhysCore.3.2.012},
}

@Article{Torre_2024_scicore,
	title={{Long-range entanglement and topological excitations}},
	author={Gianpaolo Torre and Jovan Odavić and Pierre Fromholz and Salvatore Marco Giampaolo and Fabio Franchini},
	journal={SciPost Phys. Core},
	volume={7},
	pages={050},
	year={2024},
	publisher={SciPost},
	doi={10.21468/SciPostPhysCore.7.3.050},
	url={https://scipost.org/10.21468/SciPostPhysCore.7.3.050},
}

@article{Hannes_2016_PRX,
  title = {Measurement Protocol for the Entanglement Spectrum of Cold Atoms},
  author = {Pichler, Hannes and Zhu, Guanyu and Seif, Alireza and Zoller, Peter and Hafezi, Mohammad},
  journal = {Phys. Rev. X},
  volume = {6},
  issue = {4},
  pages = {041033},
  numpages = {12},
  year = {2016},
  month = {Nov},
  publisher = {American Physical Society},
  doi = {10.1103/PhysRevX.6.041033},
  url = {https://link.aps.org/doi/10.1103/PhysRevX.6.041033}
}

@article{Kenny_2018_PRL,
  title = {Measurement of the Entanglement Spectrum of a Symmetry-Protected Topological State Using the IBM Quantum Computer},
  author = {Choo, Kenny and von Keyserlingk, Curt W. and Regnault, Nicolas and Neupert, Titus},
  journal = {Phys. Rev. Lett.},
  volume = {121},
  issue = {8},
  pages = {086808},
  numpages = {5},
  year = {2018},
  month = {Aug},
  publisher = {American Physical Society},
  doi = {10.1103/PhysRevLett.121.086808},
  url = {https://link.aps.org/doi/10.1103/PhysRevLett.121.086808}
}

@article{Vermersch_2018_PRA,
  title = {Unitary $n$-designs via random quenches in atomic Hubbard and spin models: Application to the measurement of R\'enyi entropies},
  author = {Vermersch, B. and Elben, A. and Dalmonte, M. and Cirac, J. I. and Zoller, P.},
  journal = {Phys. Rev. A},
  volume = {97},
  issue = {2},
  pages = {023604},
  numpages = {10},
  year = {2018},
  month = {Feb},
  publisher = {American Physical Society},
  doi = {10.1103/PhysRevA.97.023604},
  url = {https://link.aps.org/doi/10.1103/PhysRevA.97.023604}
}

@article{Lavasani_2021_NP,
title={Measurement-induced topological entanglement transitions in symmetric random quantum circuits},
author={Lavasani, Ali and Alavirad, Yahya and Barkeshli, Maissam},
journal={Nature Physics},
volume={17},
number={3},
pages={342--347},
year={2021},
publisher={Nature Publishing Group UK London}
}

@article{Lahtinen_2015_PRL,
  title = {Realizing All $so(N{)}_{1}$ Quantum Criticalities in Symmetry Protected Cluster Models},
  author = {Lahtinen, Ville and Ardonne, Eddy},
  journal = {Phys. Rev. Lett.},
  volume = {115},
  issue = {23},
  pages = {237203},
  numpages = {5},
  year = {2015},
  month = {Dec},
  publisher = {American Physical Society},
  doi = {10.1103/PhysRevLett.115.237203},
  url = {https://link.aps.org/doi/10.1103/PhysRevLett.115.237203}
}

@article{Doherty_2009_PRL,
  title = {Identifying Phases of Quantum Many-Body Systems That Are Universal for Quantum Computation},
  author = {Doherty, Andrew C. and Bartlett, Stephen D.},
  journal = {Phys. Rev. Lett.},
  volume = {103},
  issue = {2},
  pages = {020506},
  numpages = {4},
  year = {2009},
  month = {Jul},
  publisher = {American Physical Society},
  doi = {10.1103/PhysRevLett.103.020506},
  url = {https://link.aps.org/doi/10.1103/PhysRevLett.103.020506}
}

@article{Smacchia_2011_PRA,
  title = {Statistical mechanics of the cluster Ising model},
  author = {Smacchia, Pietro and Amico, Luigi and Facchi, Paolo and Fazio, Rosario and Florio, Giuseppe and Pascazio, Saverio and Vedral, Vlatko},
  journal = {Phys. Rev. A},
  volume = {84},
  issue = {2},
  pages = {022304},
  numpages = {12},
  year = {2011},
  month = {Aug},
  publisher = {American Physical Society},
  doi = {10.1103/PhysRevA.84.022304},
  url = {https://link.aps.org/doi/10.1103/PhysRevA.84.022304}
}

@article{Niu_2012_PRB,
  title = {Majorana zero modes in a quantum Ising chain with longer-ranged interactions},
  author = {Niu, Yuezhen and Chung, Suk Bum and Hsu, Chen-Hsuan and Mandal, Ipsita and Raghu, S. and Chakravarty, Sudip},
  journal = {Phys. Rev. B},
  volume = {85},
  issue = {3},
  pages = {035110},
  numpages = {10},
  year = {2012},
  month = {Jan},
  publisher = {American Physical Society},
  doi = {10.1103/PhysRevB.85.035110},
  url = {https://link.aps.org/doi/10.1103/PhysRevB.85.035110}
}

@article{DeGottardi_2013_PRB,
  title = {Majorana fermions in superconducting wires: Effects of long-range hopping, broken time-reversal symmetry, and potential landscapes},
  author = {DeGottardi, Wade and Thakurathi, Manisha and Vishveshwara, Smitha and Sen, Diptiman},
  journal = {Phys. Rev. B},
  volume = {88},
  issue = {16},
  pages = {165111},
  numpages = {22},
  year = {2013},
  month = {Oct},
  publisher = {American Physical Society},
  doi = {10.1103/PhysRevB.88.165111},
  url = {https://link.aps.org/doi/10.1103/PhysRevB.88.165111}
}

@article{Zeng_2016_EL,
doi = {10.1209/0295-5075/113/56001},
url = {https://doi.org/10.1209/0295-5075/113/56001},
year = {2016},
month = {mar},
publisher = {EDP Sciences, IOP Publishing and Società Italiana di Fisica},
volume = {113},
number = {5},
pages = {56001},
author = {Zeng, Bei and Zhou, D. L.},
title = {Topological and error-correcting properties for symmetry-protected topological order},
journal = {Europhysics Letters}
}

@article{Verresen_2017_PRB,
  title = {One-dimensional symmetry protected topological phases and their transitions},
  author = {Verresen, Ruben and Moessner, Roderich and Pollmann, Frank},
  journal = {Phys. Rev. B},
  volume = {96},
  issue = {16},
  pages = {165124},
  numpages = {23},
  year = {2017},
  month = {Oct},
  publisher = {American Physical Society},
  doi = {10.1103/PhysRevB.96.165124},
  url = {https://link.aps.org/doi/10.1103/PhysRevB.96.165124}
}

@article{Morral_2023_PRB,
  title = {Detecting and stabilizing measurement-induced symmetry-protected topological phases in generalized cluster models},
  author = {Morral-Yepes, Ra\'ul and Pollmann, Frank and Lovas, Izabella},
  journal = {Phys. Rev. B},
  volume = {108},
  issue = {22},
  pages = {224304},
  numpages = {14},
  year = {2023},
  month = {Dec},
  publisher = {American Physical Society},
  doi = {10.1103/PhysRevB.108.224304},
  url = {https://link.aps.org/doi/10.1103/PhysRevB.108.224304}
}

@article{Pachos_2004_PRL,
  title = {Three-Spin Interactions in Optical Lattices and Criticality in Cluster Hamiltonians},
  author = {Pachos, Jiannis K. and Plenio, Martin B.},
  journal = {Phys. Rev. Lett.},
  volume = {93},
  issue = {5},
  pages = {056402},
  numpages = {4},
  year = {2004},
  month = {Jul},
  publisher = {American Physical Society},
  doi = {10.1103/PhysRevLett.93.056402},
  url = {https://link.aps.org/doi/10.1103/PhysRevLett.93.056402}
}

@article{Montes_2012_PRE,
  title = {Phase diagram and quench dynamics of the cluster-$XY$ spin chain},
  author = {Montes, Sebasti\'an and Hamma, Alioscia},
  journal = {Phys. Rev. E},
  volume = {86},
  issue = {2},
  pages = {021101},
  numpages = {9},
  year = {2012},
  month = {Aug},
  publisher = {American Physical Society},
  doi = {10.1103/PhysRevE.86.021101},
  url = {https://link.aps.org/doi/10.1103/PhysRevE.86.021101}
}

@article{Son_2012_QIP,
  title={Topological order in 1D Cluster state protected by symmetry},
  author={Son, Wonmin and Amico, Luigi and Vedral, Vlatko},
  journal={Quantum Information Processing},
  volume={11},
  number={6},
  pages={1961--1968},
  year={2012},
  doi={10.1007/s11128-011-0346-7}
}

@article{Giampaolo_2015_PRA,
  title = {Topological and nematic ordered phases in many-body cluster-Ising models},
  author = {Giampaolo, S. M. and Hiesmayr, B. C.},
  journal = {Phys. Rev. A},
  volume = {92},
  issue = {1},
  pages = {012306},
  numpages = {9},
  year = {2015},
  month = {Jul},
  publisher = {American Physical Society},
  doi = {10.1103/PhysRevA.92.012306},
  url = {https://link.aps.org/doi/10.1103/PhysRevA.92.012306}
}

@article{Zhang_2015_PRL,
  title = {Topological Characterization of Extended Quantum Ising Models},
  author = {Zhang, G. and Song, Z.},
  journal = {Phys. Rev. Lett.},
  volume = {115},
  issue = {17},
  pages = {177204},
  numpages = {5},
  year = {2015},
  month = {Oct},
  publisher = {American Physical Society},
  doi = {10.1103/PhysRevLett.115.177204},
  url = {https://link.aps.org/doi/10.1103/PhysRevLett.115.177204}
}

@article{Li_2025_PRA,
  title = {Global phase diagram of the cluster-$XY$ spin chain with dissipation},
  author = {Li, Wei-Lin and Chen, Ying-Ao and Guo, Zheng-Xin and Yu, Xue-Jia and Li, Zhi},
  journal = {Phys. Rev. A},
  volume = {111},
  issue = {1},
  pages = {013316},
  numpages = {12},
  year = {2025},
  month = {Jan},
  publisher = {American Physical Society},
  doi = {10.1103/PhysRevA.111.013316},
  url = {https://link.aps.org/doi/10.1103/PhysRevA.111.013316}
}

@article{Ding_2019_PRE,
  title = {Phase transitions of a cluster Ising model},
  author = {Ding, Chengxiang},
  journal = {Phys. Rev. E},
  volume = {100},
  issue = {4},
  pages = {042131},
  numpages = {9},
  year = {2019},
  month = {Oct},
  publisher = {American Physical Society},
  doi = {10.1103/PhysRevE.100.042131},
  url = {https://link.aps.org/doi/10.1103/PhysRevE.100.042131}
}

@article{Guo_2022_PRA,
  title = {Emergent phase transitions in a cluster Ising model with dissipation},
  author = {Guo, Zheng-Xin and Yu, Xue-Jia and Hu, Xi-Dan and Li, Zhi},
  journal = {Phys. Rev. A},
  volume = {105},
  issue = {5},
  pages = {053311},
  numpages = {8},
  year = {2022},
  month = {May},
  publisher = {American Physical Society},
  doi = {10.1103/PhysRevA.105.053311},
  url = {https://link.aps.org/doi/10.1103/PhysRevA.105.053311}
}

@article{Yi_2025_PRA,
  title = {Continuously varying critical exponents in an exactly solvable long-range cluster $XY$ model},
  author = {Yi, Tian-Cheng and Ding, Chengxiang and Liu, Maoxin and Li, Liangsheng and You, Wen-Long},
  journal = {Phys. Rev. A},
  volume = {111},
  issue = {2},
  pages = {023307},
  numpages = {9},
  year = {2025},
  month = {Feb},
  publisher = {American Physical Society},
  doi = {10.1103/PhysRevA.111.023307},
  url = {https://link.aps.org/doi/10.1103/PhysRevA.111.023307}
}

@article{Odavic_2023_Quantum,
  doi = {10.22331/q-2023-09-15-1115},
  url = {https://doi.org/10.22331/q-2023-09-15-1115},
  title = {Random unitaries, {R}obustness, and {C}omplexity of {E}ntanglement},
  author = {Odavi{\'{c}}, J. and Torre, G. and Miji{\'{c}}, N. and Davidovi{\'{c}}, D. and Franchini, F. and Giampaolo, S. M.},
  journal = {{Quantum}},
  issn = {2521-327X},
  publisher = {{Verein zur F{\"{o}}rderung des Open Access Publizierens in den Quantenwissenschaften}},
  volume = {7},
  pages = {1115},
  month = sep,
  year = {2023}
}

@article{Yu_2025_PS,
doi = {10.1088/1402-4896/add656},
url = {https://doi.org/10.1088/1402-4896/add656},
year = {2025},
month = {may},
publisher = {IOP Publishing},
volume = {100},
number = {6},
pages = {065950},
author = {Yu, Hui and Hu, Jiangping},
title = {Measurement-driven transitions between area law phases},
journal = {Physica Scripta}
}

@article{Nie_2017_PRE,
  title = {Scaling of geometric phase versus band structure in cluster-Ising models},
  author = {Nie, Wei and Mei, Feng and Amico, Luigi and Kwek, Leong Chuan},
  journal = {Phys. Rev. E},
  volume = {96},
  issue = {2},
  pages = {020106},
  numpages = {5},
  year = {2017},
  month = {Aug},
  publisher = {American Physical Society},
  doi = {10.1103/PhysRevE.96.020106},
  url = {https://link.aps.org/doi/10.1103/PhysRevE.96.020106}
}

@article{Potirniche_2017_PRL,
  title = {Floquet Symmetry-Protected Topological Phases in Cold-Atom Systems},
  author = {Potirniche, I.-D. and Potter, A. C. and Schleier-Smith, M. and Vishwanath, A. and Yao, N. Y.},
  journal = {Phys. Rev. Lett.},
  volume = {119},
  issue = {12},
  pages = {123601},
  numpages = {6},
  year = {2017},
  month = {Sep},
  publisher = {American Physical Society},
  doi = {10.1103/PhysRevLett.119.123601},
  url = {https://link.aps.org/doi/10.1103/PhysRevLett.119.123601}
}

@article{Zonzo_2018_JSM,
doi = {10.1088/1742-5468/aac443},
url = {https://doi.org/10.1088/1742-5468/aac443},
year = {2018},
month = {jun},
publisher = {IOP Publishing and SISSA},
volume = {2018},
number = {6},
pages = {063103},
author = {Zonzo, G and Giampaolo, S M},
title = {n-cluster models in a transverse magnetic field},
journal = {Journal of Statistical Mechanics: Theory and Experiment}
}

@article{Bera_2025_arxiv,
  title={Generation of Volume-Law Entanglement by Local-Measurement-Only Quantum Dynamics},
  author={Surajit Bera and Igor V. Gornyi and Sumilan Banerjee and Yuval Gefen},
  journal={arXiv.2509.14329},
  year={2025}
}

@article{Qian_2024_PRB,
  title = {Steering-induced phase transition in measurement-only quantum circuits},
  author = {Qian, Dongheng and Wang, Jing},
  journal = {Phys. Rev. B},
  volume = {109},
  issue = {2},
  pages = {024301},
  numpages = {12},
  year = {2024},
  month = {Jan},
  publisher = {American Physical Society},
  doi = {10.1103/PhysRevB.109.024301},
  url = {https://link.aps.org/doi/10.1103/PhysRevB.109.024301}
}

@article{Ippoliti_2021_PRX,
  title = {Entanglement Phase Transitions in Measurement-Only Dynamics},
  author = {Ippoliti, Matteo and Gullans, Michael J. and Gopalakrishnan, Sarang and Huse, David A. and Khemani, Vedika},
  journal = {Phys. Rev. X},
  volume = {11},
  issue = {1},
  pages = {011030},
  numpages = {23},
  year = {2021},
  month = {Feb},
  publisher = {American Physical Society},
  doi = {10.1103/PhysRevX.11.011030},
  url = {https://link.aps.org/doi/10.1103/PhysRevX.11.011030}
}

@article{Bhattacharjee_2018_PRB,
  title = {Dynamical quantum phase transitions in extended transverse Ising models},
  author = {Bhattacharjee, Sourav and Dutta, Amit},
  journal = {Phys. Rev. B},
  volume = {97},
  issue = {13},
  pages = {134306},
  numpages = {8},
  year = {2018},
  month = {Apr},
  publisher = {American Physical Society},
  doi = {10.1103/PhysRevB.97.134306},
  url = {https://link.aps.org/doi/10.1103/PhysRevB.97.134306}
}

@article{Scaffidi_2017_PRX,
  title = {Gapless Symmetry-Protected Topological Order},
  author = {Scaffidi, Thomas and Parker, Daniel E. and Vasseur, Romain},
  journal = {Phys. Rev. X},
  volume = {7},
  issue = {4},
  pages = {041048},
  numpages = {16},
  year = {2017},
  month = {Nov},
  publisher = {American Physical Society},
  doi = {10.1103/PhysRevX.7.041048},
  url = {https://link.aps.org/doi/10.1103/PhysRevX.7.041048}
}

@article{Zhang_2018_PRL,
  title = {Characterization of Topological States via Dual Multipartite Entanglement},
  author = {Zhang, Yu-Ran and Zeng, Yu and Fan, Heng and You, J. Q. and Nori, Franco},
  journal = {Phys. Rev. Lett.},
  volume = {120},
  issue = {25},
  pages = {250501},
  numpages = {7},
  year = {2018},
  month = {Jun},
  publisher = {American Physical Society},
  doi = {10.1103/PhysRevLett.120.250501},
  url = {https://link.aps.org/doi/10.1103/PhysRevLett.120.250501}
}

@article{Wyner_1978_IC,
title = {A definition of conditional mutual information for arbitrary ensembles},
journal = {Information and Control},
volume = {38},
number = {1},
pages = {51-59},
year = {1978},
issn = {0019-9958},
doi = {https://doi.org/10.1016/S0019-9958(78)90026-8},
url = {https://www.sciencedirect.com/science/article/pii/S0019995878900268},
author = {A.D. Wyner},
}

@article{Fendley_2007_JSP,
  author  = {Fendley, Paul and Fisher, Matthew P. A. and Nayak, Chetan},
  title   = {Topological Entanglement Entropy from the Holographic Partition Function},
  journal = {J. Stat. Phys.},
  volume  = {126},
  number  = {6},
  pages   = {1111-1144},
  year    = {2007},
  month   = mar,
  doi     = {10.1007/s10955-006-9275-8},
  issn    = {1572-9613},
  url     = {https://doi.org/10.1007/s10955-006-9275-8},
}

@article{Wang_2018,
doi = {10.1209/0295-5075/124/50005},
url = {https://doi.org/10.1209/0295-5075/124/50005},
year = {2018},
month = {dec},
publisher = {EDP Sciences, IOP Publishing and Società Italiana di Fisica},
volume = {124},
number = {5},
pages = {50005},
author = {Wang, Qiang and Wang, Da and Wang, Qiang-Hua},
title = {Entanglement in a second-order topological insulator on a square lattice},
journal = {Europhys. Lett.}
}

@article{Casini_2004_PLB,
title = {A finite entanglement entropy and the c-theorem},
journal = {Phys. Lett. B},
volume = {600},
number = {1},
pages = {142-150},
year = {2004},
issn = {0370-2693},
doi = {https://doi.org/10.1016/j.physletb.2004.08.072},
url = {https://www.sciencedirect.com/science/article/pii/S037026930401264X},
author = {H. Casini and M. Huerta},
}

@article{Fabio_2022_PRB,
  title = {Fate of local order in topologically frustrated spin chains},
  author = {Mari\ifmmode \acute{c}\else \'{c}\fi{}, V. and Giampaolo, S. M. and Franchini, Fabio},
  journal = {Phys. Rev. B},
  volume = {105},
  issue = {6},
  pages = {064408},
  numpages = {16},
  year = {2022},
  month = {Feb},
  publisher = {American Physical Society},
  doi = {10.1103/PhysRevB.105.064408},
  url = {https://link.aps.org/doi/10.1103/PhysRevB.105.064408}
}

@article{data,
  author  = {Zheng, Z.-Y.},
  title   = {Dataset},
  journal = {Zenodo},
  year    = {2026},
  doi     = {https://doi.org/10.5281/zenodo.18766742}
}

\end{document}